\documentclass[         %
aps,                    %  American Physical Society
prd,                    %  Physical Review D
showpacs,               %  Displays PACS after abstract
superscriptaddress,    %  Authors' addresses linked with superscripts
nofootinbib,            %  Does not treat footnotes as references
showkeys,               %
preprintnumbers,        %
floatfix]               %  Fixes float errors
{revtex4}               %  REVTEX 4 Package used
\usepackage{graphicx,longtable,amsmath}
\numberwithin{equation}{section}
\begin{document}
%-----------------------------------------------------------------%
\title{Neutrino-$^{13}C$ Cross Sections at Supernova Neutrino Energies}
%-----------------------------------------------------------------%
%-----------------------------------------------------------------%
\author{T. Suzuki}
\email{suzuki@phys.chs.nihon-u.ac.jp}
\affiliation{Department of Physics, College of Humanities and Sciences, Nihon University, 
Sakurajosui 3-25-40, Setagaya-ku, Tokyo 156-8550, Japan}
\affiliation{National Astronomical Observatory of Japan 2-21-1 
Osawa, Mitaka, Tokyo, 181-8588, Japan}
\author{A.~B. Balantekin}
\email{baha@physics.wisc.edu}
\affiliation{Department of Physics, University of Wisconsin, Madison, WI 53706, USA}
\affiliation{National Astronomical Observatory of Japan 2-21-1 
Osawa, Mitaka, Tokyo, 181-8588, Japan}
\author{T. Kajino}
\email{kajino@nao.ac.jp}
\affiliation{National Astronomical Observatory of Japan 2-21-1 
Osawa, Mitaka, Tokyo, 181-8588, Japan}
\affiliation{Graduate School of Science, 
University of Tokyo, 7-3-1 Hongo, Bunkyo-ku, Tokyo, 113-0033, Japan}
\affiliation{School of Physics, Beihang University, Beijing 100083, China}
\author{S. Chiba}
\affiliation{Laboratory for Advanced Nuclear Energy, 
Institute of Innovative Research, 
Tokyo Institute of Technology, 
N1-9, 2-12-1 Ookayama, Meguro, Tokyo 152-8550}
\affiliation{National Astronomical Observatory of Japan 2-21-1 
Osawa, Mitaka, Tokyo, 181-8588, Japan}

%-----------------------------------------------------------------%
\date{\today}
%-----------------------------------------------------------------%
\begin{abstract}

We present neutrino capture cross sections on $^{13}$C at supernova neutrino energies, up to 50 MeV. For both charged-current and neutral-current reactions partial cross sections are calculated using statistical Hauser-Feschbach method. Coherent elastic neutrino scattering cross section for a $^{13}$C target is also provided. 

\end{abstract}
%-----------------------------------------------------------------%
\medskip
\preprint{}
\maketitle
%------------------------------------------------------------------%

\vskip 1.3cm

%%%%%%%%%%%%%%%%%%%%%%%%%%%%%%%%%%%%%%
\section{Introduction}
\label{Section: Introduction}
%%%%%%%%%%%%%%%%%%%%%%%%%%%%%%%%%%%%%%
There has been outstanding progress in all aspects of neutrino physics during the past several decades. This progress 
increasingly necessitates the availability of accurate neutrino-nucleus cross sections. The most reliable calculations of such cross sections 
at lower energies (defined as reactor, solar and supernova neutrino energies) are based on the nuclear shell model. In parallel to the rapid progress in 
neutrino physics, new shell model Hamiltonians have been recently developed.  These Hamiltonians, including the tensor forces, 
describe well nuclear spin responses and shell evolution towards the drip lines. Such a new shell model Hamiltonian for p-shell nuclei 
is the SFO (Suzuki, Fujimoto, Otsuka) Hamiltonian \cite{Suzuki:2003}. 

Many large neutrino experiments such as currently operating Daya Bay \cite{Adey:2018zwh}, RENO \cite{Bak:2018ydk}, Double Chooz \cite{Kaneda:2018nma}, BOREXINO \cite{Agostini:2018uly} experiments, under construction JUNO experiment \cite{Miramonti:2018noj}, and the proposed LENA experiment \cite{Wurm:2011zn} 
 utilize carbon-based liquid scintillators and liquid track detectors. Indeed that is one reason why the 
neutrino interactions with $^{12}$C is experimentally well studied \cite{Athanassopoulos:1997rm}.  
The natural abundance of $^{13}$C is 1.07\%, hence in a precision experiment 
neutrino interactions with $^{13}$C are no longer negligible.  
This cross section has not been directly measured, but may be accessible with a $^{13}C$-enriched target. In addition, as we discuss in Section IVb, knowing the exact value of this cross section in the Standard Model would help searches for physics beyond the Standard Model. 
In the late 1980's a $^{13}$C-enriched target was proposed as a possible solar neutrino detector \cite{Arafune:1988hx}, motivating a calculation of this reaction \cite{Fukugita:1989wv} both in the shell model (using 
Cohen-Kurath wave functions \cite{Cohen:1965qa,Millener:1975zz}) and using the effective operator method \cite{Arima:1988xa}. 
Ground-to-ground state transitions and inclusive cross sections were 
also given \cite{Pourkaviani:1991pb,Mintz:2000xv}. In a previous publication \cite{Suzuki:2012aa} 
we provided a detailed calculation of the neutrino interactions with $^{13}$C using the SFO Hamiltonian at reactor energies. 
Carbon-based scintillators can be used as supernova detectors \cite{Scholberg:2012id,Dasgupta:2011wg} and  multi-purpose neutrino observatories 
which use liquid scintillators have been proposed \cite{Wurm:2011zn}. 
Sensitivity to all flavors of neutrinos and antineutrinos in a supernova burst is the ultimate goal of a supernova neutrino detector. This goal in turn requires knowing as many as possible cross sections for various potential detectors. 
The goal of the present paper 
is to extend our previous calculations of neutrino-$^{13}$C cross sections to higher energies, up to about 50 MeV, relevant to supernova observations and experiments at the spallation neutron sources.  

\section{Calculations with the SFO Hamiltonian}

Starting from the Cohen-Kurath (CK) \cite{Cohen:1965qa} and the Millener-Kurath (MK) \cite{Millener:1975zz} Hamiltonians, a new shell-model Hamiltonian for the p-sd shell has been constructed. 
The CK Hamiltonians are phenomenological effective interactions for the p-shell obtained by fitting experimental low-lying energy levels. In one version, (8-16)2BE, single-particle energies for $p_{1/2}$ and $p_{3/2}$ orbits and two-body matrix elements are determined by using 35 energy data for nuclei with mass numbers A =8-16. 
The MK Hamiltonian is an effective particle-hole interaction for p-sd cross shell obtained by fitting to energy levels of non-normal parity states of nuclei with A =11-16. The strength parameters of central, tensor and two-body spin-orbit potentials are determined. 
The new Hamiltonian for the p-sd shell, SFO \cite{Suzuki:2003}, consists of p-shell, sd-shell and p-sd shell parts, where $\langle p, sd \mid V \mid p, sd\rangle$ part is taken from the MK Hamiltonian while $\langle p^2 \mid V \mid (sd)^2 \rangle$ and 
%$\langle (sd)^2 \mid V \mid (sd)^2 \rangle$ 
sd-shell parts are from Kuo's G-matrix calculation \cite{Kuo:1967}. 
The p-shell part is modified from (8-16)2BME by enhancing the magnitude of the monopole terms of the matrix elements for     
%Starting from the interactions of Refs. \cite{Cohen:1965qa} and \cite{Millener:1975zz}, the SFO Hamiltonian is constructed by enhancing monopole terms of the matrix elements for 
the $p_{1/2}-p_{3/2}$ orbits with zero isospin as well as the single-particle energy gap between the $p_{1/2}$ and $p_{3/2}$ orbits.
The monopole term is changed by -2.14 MeV and the gap is enhanced from 0.14 MeV to 3.92 MeV. 

Using a configuration space including up to $2\hbar \omega$ excitations and a small 
(five percent) quenching of the axial-vector coupling constant and spin g factor, this Hamiltonian considerably improves the magnetic properties 
of the p-shell nuclei as compared with the earlier treatments using CK interactions. For example, the Gamow-Teller strength in $^{12}$C is 
well described by the SFO Hamiltonian, resulting in a good agreement with the experimental measurements \cite{Suzuki:2006qd}. 
The enhancement of the magnitude of the monopole terms in the spin-isospin flip channel leads to more admixtures of $p_{1/2}$ and $p_{3/2}$ shell components in the wave functions and reduction of the $B(GT)$ strength.
The enhancement of the single-particle energy gap, on the other hand, leads to less admixtures of $p_{1/2}$ and $p_{3/2}$ shell components and enhancement of the $B(GT)$ value.
Spin and magnetic properties of nuclei are sensitive to the balance between the monopole terms and the single-particle energy gap.  Their proper choice is important for the description of the spin degree's of freedom in nuclei. 
The SFO Hamiltonian contains the proper tensor components consistent with the sign rules for the monopole-tensor terms \cite{Otsuka:2005zz}, 
that is, attractive for j$_{\rangle}$-j$_{\langle}$ orbits (j$_{\rangle}$=$\ell$+1/2, j$_{\langle}$ = $\ell$-1/2) but repulsive for j$_{\rangle}$-j$_{\rangle}$ or j$_{\langle}$-j$_{\langle}$ orbits. 
The spin-isospin dependent part of the SFO interaction is strong enough to be consistent with this sign rule, while the CK interaction contains the tensor components with opposite signs to the proper ones due to its weak monopole terms in the spin-isospin flip channel \cite{Suzuki:2006qd}.
The description of spin and magnetic properties of p-shell nuclei is thus improved for the SFO Hamiltonian compared with the CK Hamiltonians. 
The SFO interaction is found to reproduce well the exclusive cross section $^{12}$C ($\nu_e$, e$^{-}$) $^{12}$N (1$^{+}_{g.s.}$) as well as charged- and neutral-current inclusive reaction cross sections on $^{12}$C induced by pion DAR (decay-at-rest) neutrinos within error bars \cite{Suzuki:2006qd,Suzuki:2013}. 
In particular in Fig. 4 of Ref. \cite{Suzuki:2013}, the exclusive cross section 
measured (Fig. 14 of Ref. \cite{Athanassopoulos:1997rm}), which is not folded over the DAR 
spectrum, is shown to be reproduced by SFO quite well. 
Another successful example of SFO interaction is the Gamow-Teller strength in $^{14}$C \cite{Suzuki:2003}.
The $B(GT)$ strength is found to be almost vanished for the SFO interaction, where the proper inclusion of the tensor components in the interaction is essential.         
Note that the calculations with the SFO Hamiltonian are carried out in the $p$-$sd$ shell including up to 
2$\hbar\omega$ excitations with $g_{A}^{\rm eff}$/$g_A$ = 0.95, whereas those
for the comparison CK Hamiltonian are obtained within the $p$-shell with a larger quenching factor,
$g_{A}^{\rm eff}$/$g_A$ = 0.69 \cite{Fukugita:1989wv}, which is adjusted to reproduce 
the experimental $B(GT)$ values of $^{13}$N($\beta^{+}$)$^{13}$C, 
$^{15}$O($\beta^{+}$)$^{15}$N and $^{11}$C($\beta^{+}$)$^{11}$B. For the charged-current neutrino scattering on $^{13}$C the SFO Hamiltonian 
predicts an enhancement of the Gamow-Teller strength for the transition to the 3.50 MeV $3/2^-$ state in $^{13}$N as compared with the prediction of the 
CK Hamiltonian \cite{Suzuki:2012aa}. 

In our calculations neutrino-nucleus reaction cross sections are evaluated using the multipole expansion of the weak hadronic currents as described in 
Ref.  \cite{Walecka:1975}. To calculate partial cross sections for the photon and particle emission channels statistical Hauser-Feschbach method is used. 
Branching ratios from each excited level are evaluated by taking into account single- and multiparticle decay channels involving neutron, proton, deuteron, $\alpha$, $^{3}$He, $^{3}$H and $\gamma$. All the levels obtained in the present shell-model calculations are adopted as levels in the decaying and daughter nuclei with specific isospin assignments. 
The particle transmission coefficients are calculated by the optical model \cite{Walter:1986,Avrigeanu:1994}. The $\gamma$ transmission coefficients are calculated with the Brink formula. The E1 (elecric dipole) and M1 (magnetic dipole) parameters are taken from RIPL-2 database \cite{Belgya:2006}. The $\gamma$ cascade in the initial excited nuclei and subsequent decays are fully considered.

\section{Charged-current Cross Sections}

\begin{figure}[t]
\begin{center}
\includegraphics[scale=0.45]{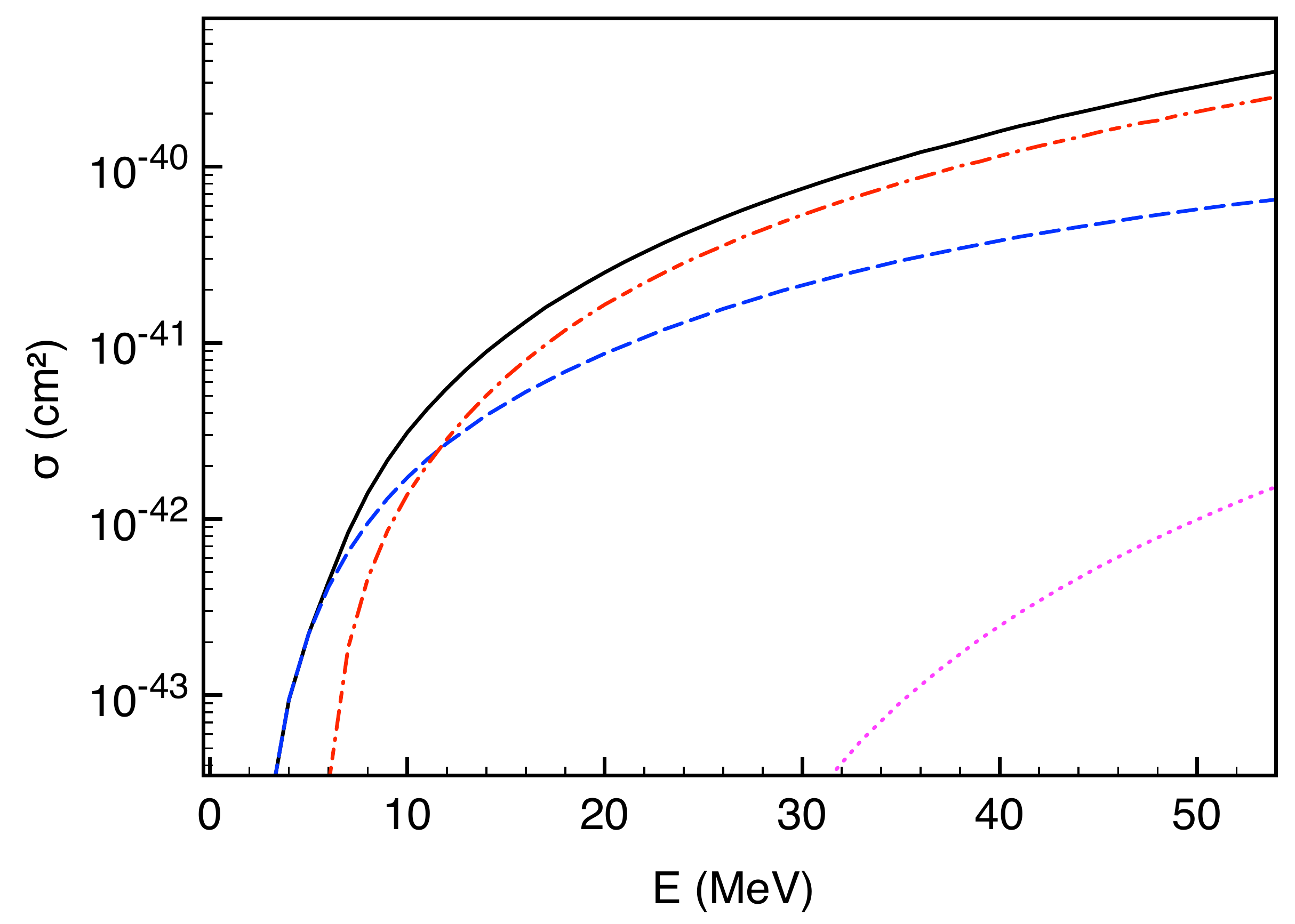}
\caption{Total charged current cross section for the $\nu_e + ^{13}$C reaction (solid line) as a function of the incoming neutrino energy. Also shown are the two major contributions to the total cross section: The proton emission cross section of Eq. (\ref{13CCCprotonat}) (dot-dashed line) and transition to the ground state of $^{13}$N (Eq. (\ref{13CCCni}), 
dashed line). Also shown the neutron emission cross section (Eq. (\ref{CCneutron}), dotted line). To avoid cluttering of the figure several other contributions the same order as the neutron emission cross section are not shown.}
\label{CCcomp}
\end{center}
\end{figure}

We present the total charged-current cross sections for $\nu_e - ^{13}$C scattering in Table I. The cross sections given in this table include all possible channels. 
\begin{table}[t]
  \centering
  \begin{tabular}{@{} cccccccc @{}}
    \hline
    $E_{\nu}$(MeV) & $\sigma$ (cm$^2$) &$E_{\nu}$(MeV) & $\sigma$ (cm$^2$)& $E_{\nu}$(MeV) & $\sigma$ (cm$^2$) 
    & $E_{\nu}$(MeV) & $\sigma$ (cm$^2$)\\ 
    \hline
3& $2.21 \times 10^{-44}$& 16 & $1.32 \times 10^{-41}$ & 29 & $6.87 \times 10^{-41}$&42&$1.80 \times 10^{-40}$\\
4& $9.39 \times 10^{-44}$& 17& $1.59 \times 10^{-41}$ & 30&$ 7.50 \times 10^{-41}$&43&$1.92 \times 10^{-40}$ \\
    5& $2.23\times 10^{-43}$ & 18 & $ 1.86 \times 10^{-41}$ &31 &$8.18 \times 10^{-41}$ &44&$2.03 \times 10^{-40}$ \\ 
    6 & $4.39 \times 10^{-43}$  & 19 & $2.17 \times 10^{-41}$ & 32& $8.89 \times 10^{-41}$&45&$2.15 \times 10^{-40}$ \\ 
    7 & $8.38 \times 10^{-43}$& 20& $2.51 \times 10^{-41}$ & 33 & $9.62 \times 10^{-41}$&46&$2.28 \times 10^{-40}$\\ 
    8 & $1.41 \times 10^{-42}$& 21 & $2.88 \times 10^{-41}$ & 34 &$1.04 \times 10^{-40}$&47&$2.41 \times 10^{-40}$\\
9& $2.16 \times 10^{-42}$ & 22 & 3.27 $ \times 10^{-41}$ & 35 & $1.12 \times 10^{-40}$&48&$2.56 \times 10^{-40}$\\
10&$3.10 \times 10^{-42}$ & 23&$3.70 \times 10^{-41}$& 36 & $1.21 \times 10^{-40}$&49&$2.70 \times 10^{-40}$\\
11&$4.20 \times 10^{-42}$&24 &$4.15\times 10^{-41}$& 37 & $1.29 \times 10^{-40}$&50&$2.84 \times 10^{-40}$\\
12& $5.52 \times 10^{-42}$& 25 &$4.62 \times 10^{-41}$ &38& $1.38 \times 10^{-40}$&51&$2.99 \times 10^{-40}$\\
 13& $7.09 \times 10^{-42}$&26&$5.14\times10^{-41}$&39&$1.48 \times 10^{-40}$&52&$3.15 \times 10^{-40}$\\
 14& $8.90 \times 10^{-42}$&27&$5.69 \times 10^{-41}$&40&$1.59 \times 10^{-40}$&53&$3.31 \times 10^{-40}$\\
 15& $1.09 \times 10^{-41}$&28&$6.26 \times 10^{-41}$&41&$1.70\times 10^{-40}$&54&$3.47 \times 10^{-40}$\\
 
    \hline
  \end{tabular}
  \caption{Total cross section for the charged current reaction $^{13}$C ($\nu_e, e^-$X)  with all channels included.}
  \label{tab:label}
\end{table}
The most dominant contribution to the total cross section comes from the proton knock-out reaction 
\begin{equation}
\label{13CCCprotonat}
\nu_e + ^{13}{\rm C} \rightarrow ^{12}{\rm C} + e^- + p .
\end{equation}
This cross section is given in Table II. This partial cross section can be observed if the scintillator  is capable of pulse-shape discrimination. 
The second dominant contribution comes from the reaction 
\begin{equation}
\label{13CCCni}
\nu_e + ^{13}{\rm C} \rightarrow ^{13}{\rm N \> (ground \> state)} + e^- .
\end{equation}
Of significant interest is the neutron emission cross section
\begin{equation}
\label{CCneutron}
\nu_e + ^{13}{\rm C} \rightarrow ^{12}{\rm N \> (ground \> state)} + e^- + n . 
\end{equation}
These neutrons thermalize and capture abundant protons typically present in the scintillators, giving the characteristic signature of 2.2 MeV photons. However the cross section for 
the reaction in Eq. (\ref{CCneutron}) is much smaller than the two dominant contributions mentioned above as depicted in Fig. \ref{CCcomp}. 
(Note that to avoid cluttering several other contributions which are the same order as the neutron emission cross section are not shown in this figure. These contributions come from emissions of photons, proton pairs, proton-neutron pairs, one proton and one alpha pairs or one proton along with a pair of alpha particles, but they are too small to be directly detectable). 
\begin{table}[b]
  \centering
  \begin{tabular}{@{} cccccccc @{}}
    \hline
    $E_{\nu}$(MeV) & $\sigma$ (cm$^2$) &$E_{\nu}$(MeV) & $\sigma$ (cm$^2$)& $E_{\nu}$(MeV) & $\sigma$ (cm$^2$) 
    & $E_{\nu}$(MeV) & $\sigma$ (cm$^2$)\\ 
    \hline
    5& $1.85\times 10^{-46}$ & 17 & $ 9.76 \times 10^{-42}$ &29 &$4.85 \times 10^{-41}$ &41&$1.23 \times 10^{-40}$ \\ 
    6 & $3.05 \times 10^{-44}$  & 18& $1.18 \times 10^{-41}$ & 30& $5.32 \times 10^{-41}$&42&$1.31 \times 10^{-40}$ \\ 
    7 & $1.86 \times 10^{-43}$& 19& $1.40 \times 10^{-41}$ & 31 & $5.82 \times 10^{-41}$&43&$1.39 \times 10^{-40}$\\ 
    8 & $4.62 \times 10^{-43}$& 20 & $1.65 \times 10^{-41}$ & 32 &$6.35 \times 10^{-41}$&44&$1.47 \times 10^{-40}$\\
9& $8.60 \times 10^{-43}$ & 21 & 1.88 $ \times 10^{-41}$ & 33 & $6.90 \times 10^{-41}$&45&$1.57 \times 10^{-40}$\\
10&$1.14 \times 10^{-42}$ & 22&$2.19 \times 10^{-41}$& 34 & $7.47 \times 10^{-41}$&46&$1.66 \times 10^{-40}$\\
11&$2.03 \times 10^{-42}$&23 &$2.50\times 10^{-41}$& 35 & $8.09 \times 10^{-41}$&47&$1.76 \times 10^{-40}$\\
12& $2.84 \times 10^{-42}$& 24 &$2.84 \times 10^{-41}$ &36& $8.71 \times 10^{-41}$&48&$1.83 \times 10^{-40}$\\
 13& $3.84 \times 10^{-42}$&25&$3.19\times10^{-41}$&37&$9.37 \times 10^{-41}$&49&$1.95 \times 10^{-40}$\\
 14& $5.02 \times 10^{-42}$&26&$3.56 \times 10^{-41}$&38&$1.01 \times 10^{-40}$&50&$2.05 \times 10^{-40}$\\
 15& $6.40 \times 10^{-42}$&27&$3.99 \times 10^{-41}$&39&$1.07\times 10^{-40}$&51&$2.16 \times 10^{-40}$\\
  16& $7.98 \times 10^{-42}$&28&$4.39 \times 10^{-41}$&40&$1.15\times 10^{-40}$&52&$2.26 \times 10^{-40}$\\
    \hline
  \end{tabular}
  \caption{Cross section for the charged current reaction $^{13}$C ($\nu_e, e^- p$) $^{12}$C.}
  \label{protonemission}
\end{table}

\begin{figure}[b]
\begin{center}
\includegraphics[scale=0.4]{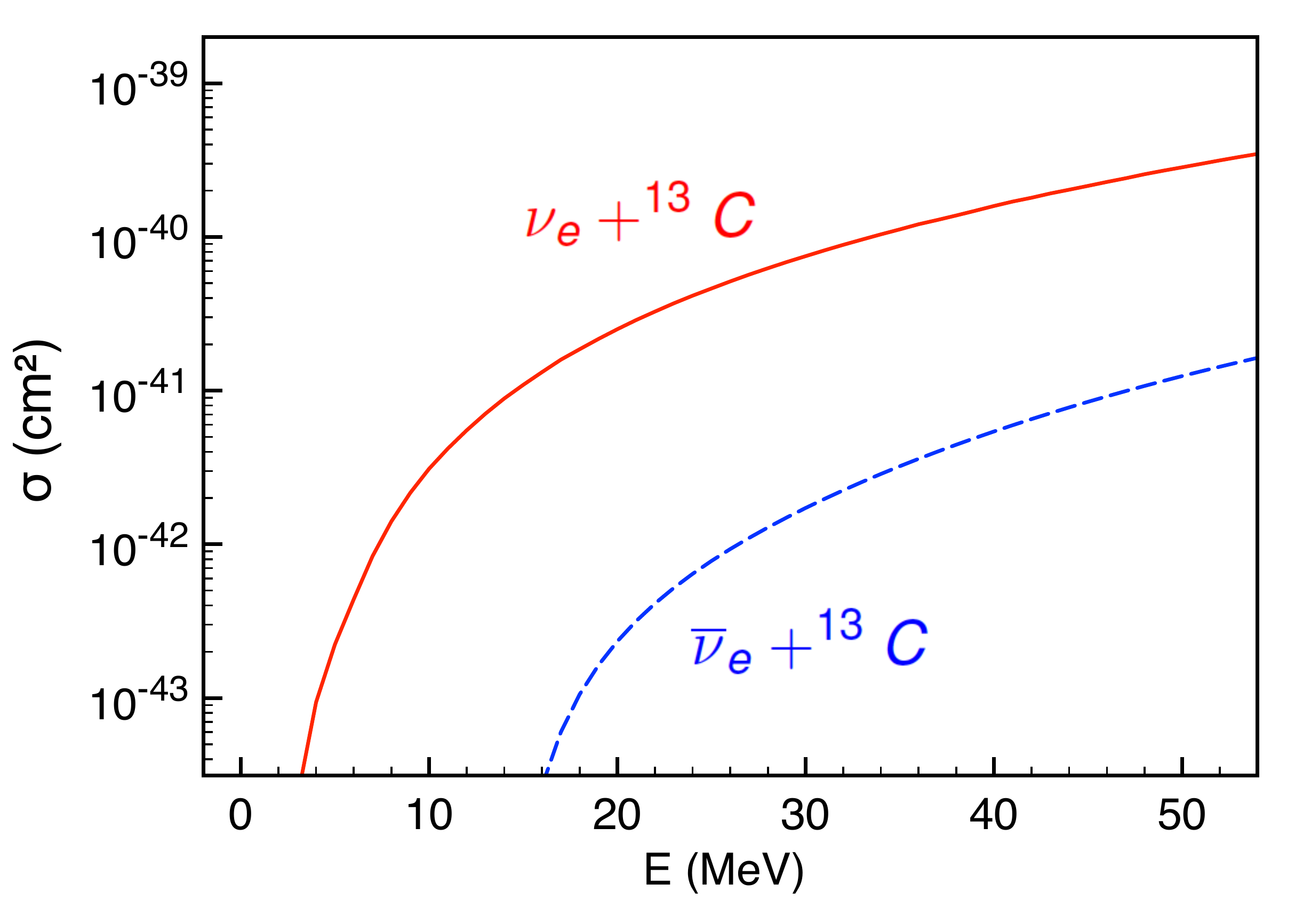}
\caption{Total charged current cross sections for the $\nu_e + ^{13}$C reaction (solid line) and  $\overline{\nu}_e + ^{13}$C reaction (dashed line) as a function of the incoming neutrino energy.}
\label{CCcompe+e-}
\end{center}
\end{figure}

The charged-current electron antineutrino capture cross section on $^{13}$C has a relatively high energy threshold ($\sim 16$ MeV). In addition it is much smaller than the charged-current electron neutrino cross section on $^{13}$C. We compare those two cross section is Fig. \ref{CCcompe+e-}.
$^{12}$C also has a relatively high energy threshold for both electron neutrinos and electron antineutrinos.  
Hence a scintillator containing $^{12}$C, $^{13}$C and protons is sensitive to electron neutrinos only via their capture on $^{13}$C and can be used to search for new physics producing electron neutrinos in reactor flux. This is why such scintillators were also proposed as solar neutrino detectors \cite{Fukugita:1989wv,Ianni:2005ki}. It should also be mentioned that neutrinos coming from spallation neutron sources contain electron neutrinos in addition to muon neutrinos and antineutrinos; however electron antineutrinos are absent from their spectra \cite{Bolozdynya:2012xv}.

\section{Neutral Current Cross Sections}

\subsection{Neutrino Coherent Scattering}

Neutrino-nucleus coherent elastic scattering cross section is given by \cite{Freedman:1977xn,Drukier:1983gj}
\begin{equation}
\label{nucoherent}
\frac{d \sigma}{dT} (E,T) = \frac{G_F^2}{8 \pi} M \left[ 2 - \frac{2T}{T_{\rm max}} +  \left( \frac{T}{E} \right)^2  \right] Q_W^2 \left[ F(Q^2)\right]^2  ,
\end{equation}
where $T$ is the recoil energy of the nucleus, $E$ is the energy of the incoming neutrino, $M$ is the mass of the target nucleus, $Q^2$ is the momentum transfer, $T_{\rm max}$ is the maximum nuclear recoil energy  
\begin{equation}
T_{\rm max} = \frac{2E^2}{2E +M}, 
\end{equation}
and  
\begin{equation}
Q_W = N -(1-4 \sin^2\theta_W)Z \nonumber 
\end{equation}
is the weak charge of the nucleus. The form factor
\begin{equation}
\label{formfactor}
F(Q^2) = \frac{1}{Q_W} \int dr \> r^2\frac{\sin (Qr)}{Qr} [ \rho_n(r) - (1-4 \sin^2\theta_W) \rho_p(r) ]
\end{equation}
corrects for contributions to scattering that are not completely coherent as $E$ gets large. In this expression $\rho_n$ and $\rho_p$ are the neutron and proton density distributions in the nucleus, respectively. In writing Eq. (\ref{formfactor}) nuclei are assumed to be at least nearly spherically symmetric. This reaction was experimentally observed for the first time only recently using a CsI scintillator \cite{Akimov:2017ade}.

\begin{figure}[t]
\begin{center}
\includegraphics[scale=0.4]{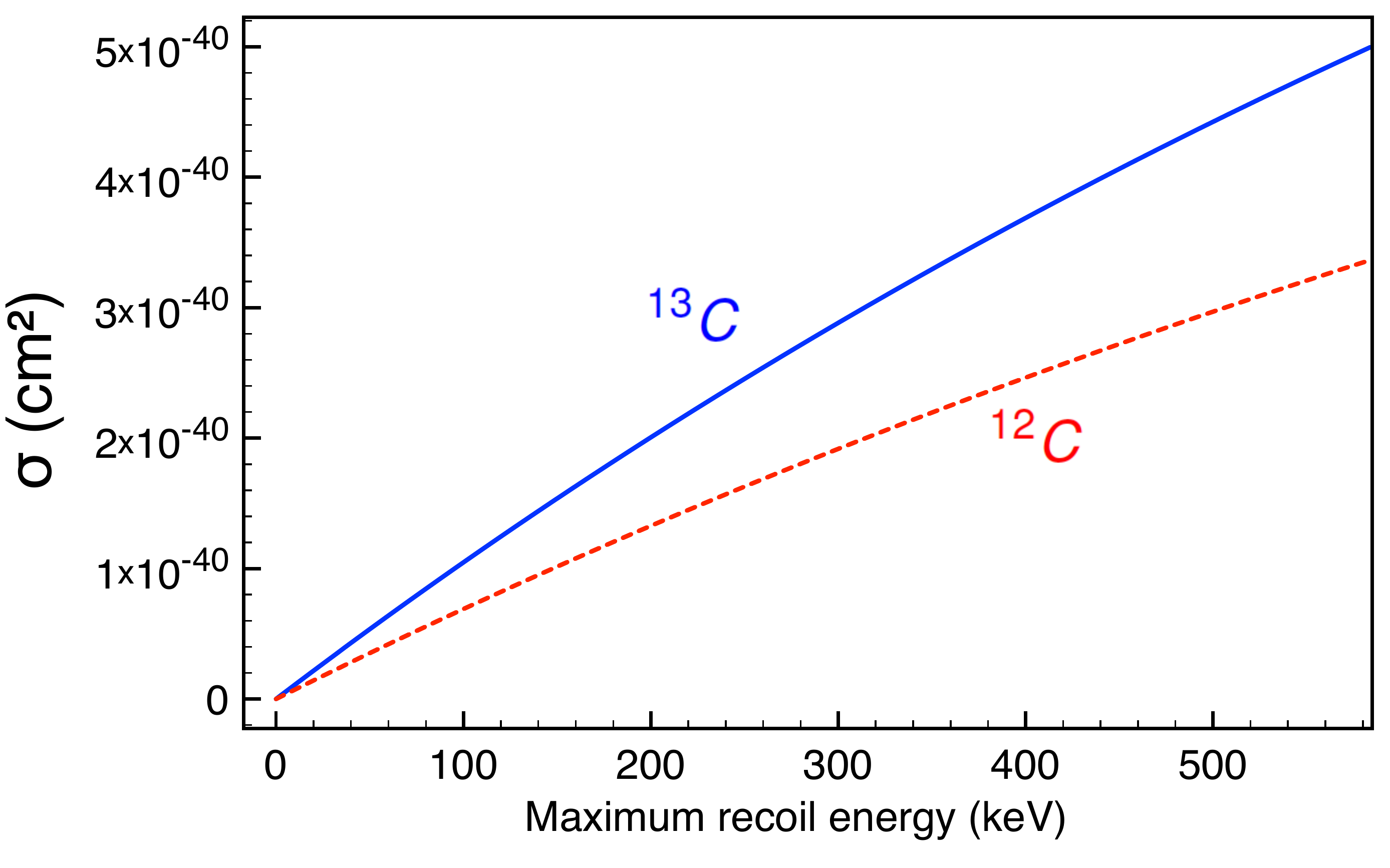}
\caption{Neutrino elastic scattering cross sections on $^{13}$C (solid line) and $^{12}$C (dashed line) as a function of the {\em maximum} nuclear recoil energy. The $^{12}$C cross section is taken from Ref. \cite{Yoshida:2008zb}.}
\label{figcf}
\end{center}
\end{figure}

The total elastic scattering cross section is given by the expression 
\begin{equation}
\sigma (E) = \int_0^{T_{\rm max}} dT \frac{d \sigma}{dT} (E, T) .
\end{equation}
We present the total elastic scattering cross section for $^{13}$C in Table \ref{tab:label}. 
\begin{table}[t]
  \centering
  \begin{tabular}{@{} cccccccc @{}}
    \hline
    $E_{\nu}$(MeV) & $\sigma$ (cm$^2$) &$E_{\nu}$(MeV) & $\sigma$ (cm$^2$)& $E_{\nu}$(MeV) & $\sigma$ (cm$^2$) 
    & $E_{\nu}$(MeV) & $\sigma$ (cm$^2$)\\ 
    \hline 
 1& $1.47 \times 10^{-43}$ & 15  &$3.96 \times 10^{-41}$    & 29& $1.41 \times 10^{-40}$   &43 & $2.89 \times 10^{-40}$  \\
2& $6.77 \times 10^{-43}$& 16 & $4.49 \times 10^{-41}$ & 30 & $1.51 \times 10^{-40}$&44&$3.01 \times 10^{-40}$\\
3& $1.58 \times 10^{-42}$& 17& $5.06 \times 10^{-41}$ & 31&$ 1.60 \times 10^{-40}$&45&$3.13 \times 10^{-40}$ \\
    4& $2.83\times 10^{-42}$ & 18 & $ 5.66 \times 10^{-41}$ &32 &$1.70 \times 10^{-40}$ &46&$3.25 \times 10^{-40}$ \\ 
    5 & $4.45 \times 10^{-42}$  & 19 & $6.28 \times 10^{-41}$ & 33& $1.80 \times 10^{-40}$&47&$3.37 \times 10^{-40}$ \\ 
    6 & $6.39 \times 10^{-42}$& 20& $6.94 \times 10^{-41}$ & 34 & $1.90 \times 10^{-40}$&48&$3.49 \times 10^{-40}$\\ 
    7 & $8.69 \times 10^{-42}$& 21 & $7.63 \times 10^{-41}$ & 35 &$2.00 \times 10^{-40}$&49&$3.61 \times 10^{-40}$\\
8& $1.14 \times 10^{-41}$ & 22 & 8.35 $ \times 10^{-41}$ & 36 & $2.11 \times 10^{-40}$&50&$3.74 \times 10^{-40}$\\
9&$1.44 \times 10^{-41}$ & 23&$9.09 \times 10^{-41}$& 37 & $2.21 \times 10^{-40}$&51&$3.86 \times 10^{-40}$\\
10&$1.77 \times 10^{-41}$&24 &$9.87\times 10^{-41}$& 38 & $2.32 \times 10^{-40}$&52&$3.98 \times 10^{-40}$\\
11& $2.14 \times 10^{-41}$& 25 &$1.07 \times 10^{-40}$ &39& $2.43 \times 10^{-40}$&53&$4.11 \times 10^{-40}$\\
 12& $2.54 \times 10^{-41}$&26&$1.15\times10^{-40}$&40&$2.55 \times 10^{-40}$&54&$4.24 \times 10^{-40}$\\
 13& $2.98 \times 10^{-41}$&27&$1.24 \times 10^{-40}$&41&$2.66 \times 10^{-40}$&55&$4.49 \times 10^{-40}$\\
 14& $3.45 \times 10^{-41}$&28&$1.32 \times 10^{-40}$&42&$2.77\times 10^{-40}$&56&$4.62 \times 10^{-40}$\\
 
    \hline
  \end{tabular}
  \caption{Total cross section for neutrino-$^{13}$C elastic scattering.}
  \label{tab:label}
\end{table}
Note that the nuclear recoil energies are very small. However, due to the extra neutron, neutrino elastic scattering from $^{13}$C is significantly more than the 
similar reaction of $^{12}$C. To illustrate this we compare the elastic scattering cross sections on $^{13}$C and $^{12}$C in Figure \ref{figcf} as a function of the 
maximum nuclear recoil energy. In this figure the coherent scattering cross section for $^{12}$C is taken from Ref. \cite{Yoshida:2008zb}. 
As this figure illustrates even a single extra neutron appreciably increases the coherent scattering cross section. 

\begin{figure}[t]
\begin{center}
\includegraphics[scale=0.4]{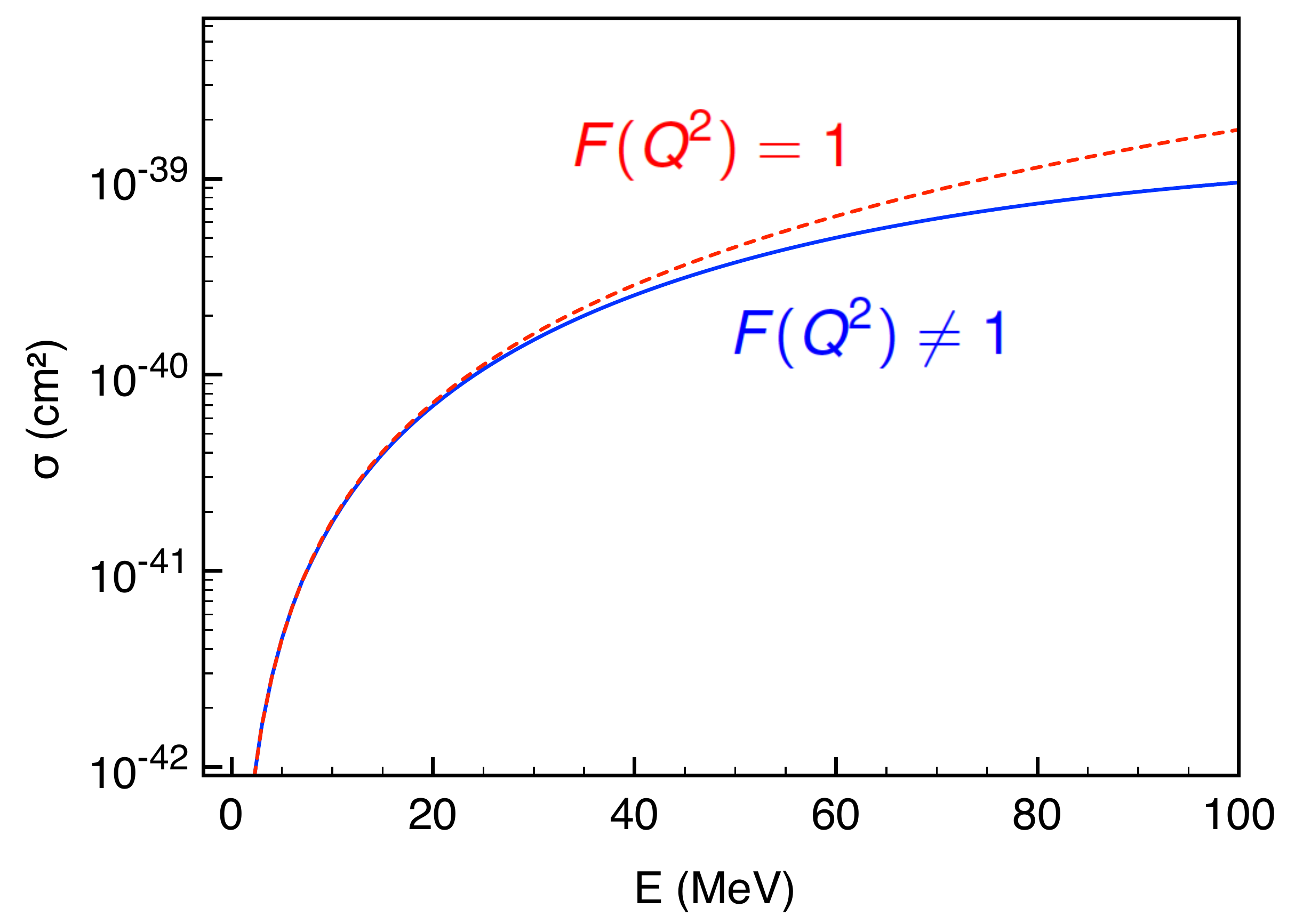}
\caption{Total neutrino elastic scattering cross sections on $^{13}$C (solid line). The prediction of Eq.(\ref{nucoherent}) with $F(Q^2)=1$ is also shown (dashed line).}
\label{figcfd}
\end{center}
\end{figure}

In Ref. \cite{Patton:2012jr} an expansion of the form factor in Eq. (\ref{formfactor}) was given in powers of the momentum transfer, $Q^2$. Since the proton contribution 
to the form factor is exceedingly small, neutrino coherent scattering primarily probes the neutron distribution. In the leading order the momentum transfer is given by 
\begin{equation}
Q^2 \sim 2MT. 
\end{equation}
For lighter nuclei an expansion of the form factor 
can be written as
\begin{equation}
\label{ffexpansion} 
F(Q^2) = 1 + \eta_2 Q^2 + \eta_4 Q^4 +  \cdots ,
\end{equation}
where the coefficients $\eta_i$ carry information about the neutron distribution. It can be shown that keeping terms up to and including $Q^4$ in the expansion of Eq. (\ref{ffexpansion}) 
contributes terms up to and including $E^6$ in the total cross section. This feature can be observed by writing the total cross section as an expansion: 
\begin{equation}
\sigma (E) = \frac{G_F^2}{8 \pi}   Q_W^2 \left[ x + \frac{4}{3} \eta_2 x^2 + \frac{1}{3} \left( 2 ( \eta_2^2 + 2 \eta_4) x^3 + \frac{x^3}{(ME)^2} \right)+ \cdots  \right]  
\end{equation}
where $x=M T_{\rm max}$. The following expansion of this expression may be more useful in applications: 
\begin{eqnarray}
\sigma (E) = \frac{G_F^2}{4\pi} Q_W^2 E^2  \left[ \left( 1 + \frac{8}{3} \eta_2 E^2 + \frac{8}{3} (\eta_2^2 + 2 \eta_4) E^4 + \cdots \right) \right. 
 \nonumber \\
\left.  - \frac{2}{M} \left( E + \frac{16}{3} \eta_2 E^3 + \frac{24}{3} (\eta_2^2 + 2 \eta_4) E^5 + \cdots \right) + \cdots \right] .
\end{eqnarray}
For $^{13}$C with only seven neutrons the contribution of even $\eta_4$ term is exceedingly small. In Fig. \ref{figcfd} we give a comparison of 
the total neutrino elastic scattering cross sections on $^{13}$C and the prediction of Eq.(\ref{nucoherent}) with $F(Q^2)=1$. Clearly nuclear structure effects decrease the cross section from the value given in the limit where deviations from $F(Q^2)=1$ are ignored. It is important to include such effects when one is exploring the decrease of the elastic scattering cross section due to other physical effects such as possible production of sterile neutrinos. One should also mention that, since oscillations experiments established the non-zero values of the neutrino masses, dipole moments of neutrinos do not vanish, very small in the Standard Model, but may receive contributions from physics beyond the Standard Model. Such magnetic moments would provide an electromagnetic contribution to the coherent scattering, but that contribution is expected to be exceedingly small.  

\subsection{Neutron knock-out}

\begin{figure}[b]
\begin{center}
\includegraphics[scale=0.4]{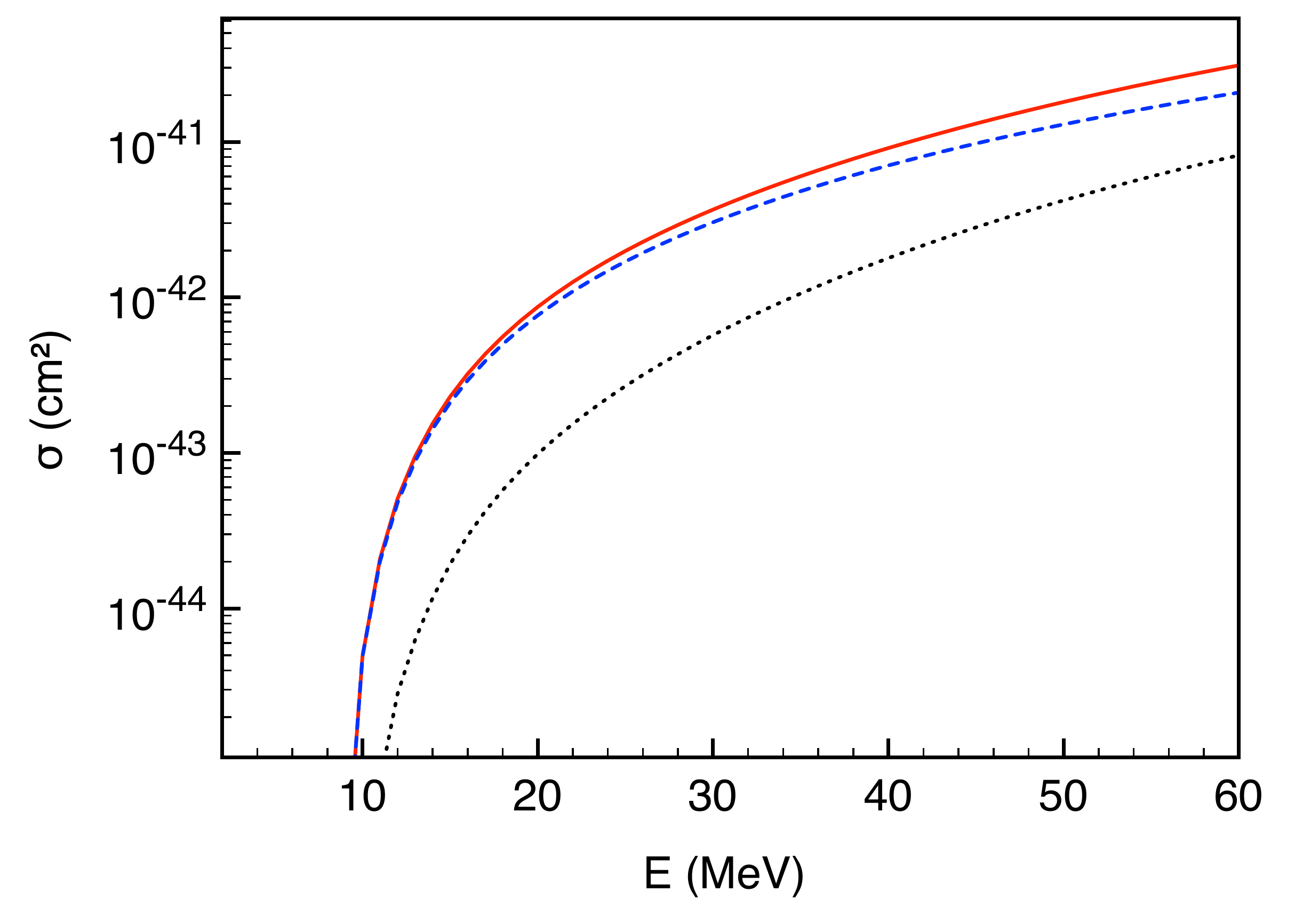}
\caption{Neutron emission cross sections for neutral-current neutrino scattering on $^{13}$C.  The total neutron emission cross section, $\nu + ^{13}C \rightarrow \nu + n + {\rm all \> final \> states}$ (solid line), cross section for transition to the  $0^+$ ground state in $^{12}$C, $\nu + ^{13}C \rightarrow \nu + n + ^{12}C ({\rm g.s.})$ (dashed line), and  the cross section for transition to the  $2^+$ state in $^{12}$C at 4.44 MeV followed by decay into the $^{12}$C ground state, $\nu + ^{13}C \rightarrow \nu + n + ^{12}C (2^+) \rightarrow \nu + n +  ^{12}C ({\rm g.s.}) + \gamma (4.44 {\rm \> MeV})$ (dotted line) are shown. }
\label{NCnemit}
\end{center}
\end{figure}

One interesting aspect of the neutral-current scattering  of neutrinos and antineutrinos on $^{13}$C is the possibility of one neutron emission in the final state. These neutrons are detectable via the 2.2 MeV photons and can provide a useful signal. In these cross sections there are contributions from interference between vector and axial-vector currents, which have different signs between left-handed neutrinos and right-handed anti-neutrinos. This feature leads to a small difference between ($\nu, \nu'$) and ($\overline{\nu}, \overline{\nu}'$) cross sections.
The total neutron emission cross sections are given in Table \ref{nuandnubarNC}.
\begin{table}[t]
  \centering
  \begin{tabular}{@{} cccccc @{}}
    \hline
     E (MeV) & $\sigma (\nu +^{13}$C) (cm$^2$) &  $\sigma (\overline{\nu} +^{13}$C) (cm$^2$) &   E (MeV) & $\sigma (\nu +^{13}$C) (cm$^2$) &  $\sigma (\overline{\nu} +^{13}$C) (cm$^2$) \\
    \hline 
6.00 & 0.161144$\times 10^{-55}$ & 0.616611$\times 10^{-55}$   &  34.00 &0.548597$\times 10^{-41}$ &0.464032$\times 10^{-41}$ \\
    7.00 & 0.498686$\times 10^{-48}$ &0.808538$\times 10^{-49}$ &  35.00 &0.601253$\times 10^{-41}$ &0.505865$\times 10^{-41}$ \\
    8.00 & 0.170467$\times 10^{-46}$ &0.418146$\times 10^{-47}$ & 36.00 &0.656985$\times 10^{-41}$ &0.549798$\times 10^{-41}$ \\
    9.00 & 0.144971$\times 10^{-45}$ &0.863204$\times 10^{-46}$ &  37.00 &0.715877$\times 10^{-41}$ &0.595865$\times 10^{-41}$ \\ 
   10.00 & 0.492129$\times 10^{-44}$ &0.461908$\times 10^{-44}$ & 38.00 &0.778015$\times 10^{-41}$ &0.644095$\times 10^{-41}$ \\
   11.00 & 0.208600$\times 10^{-43}$ &0.197935$\times 10^{-43}$ & 39.00 &0.843485$\times 10^{-41}$ &0.694513$\times 10^{-41}$ \\
   12.00 & 0.506791$\times 10^{-43}$ &0.480249$\times 10^{-43}$ & 40.00 &0.912365$\times 10^{-41}$ &0.747139$\times 10^{-41}$ \\
   13.00 &0.947193$\times 10^{-43}$ & 0.893967$\times 10^{-43}$ & 41.00& 0.984743$\times 10^{-41}$ &0.802002$\times 10^{-41}$ \\
   14.00 & 0.154234$\times 10^{-42}$ &0.144903$\times 10^{-42}$  & 42.00 &0.106069$\times 10^{-40}$ & 0.859116$\times 10^{-41}$ \\
   15.00 &0.229880$\times 10^{-42}$ &0.214934$\times 10^{-42}$  &  43.00 &0.114029$\times 10^{-40}$ &0.918500$\times 10^{-41}$\\
   16.00 & 0.322241$\times 10^{-42}$ &0.299761$\times 10^{-42}$ &  44.00 &0.122361$\times 10^{-40}$ &0.980158$\times 10^{-41}$ \\
   17.00 & 0.431917$\times 10^{-42}$ &0.399709$\times 10^{-42}$ & 45.00 &0.131073$\times 10^{-40}$ &0.104410$\times 10^{-40}$ \\
   18.00 &0.559371$\times 10^{-42}$ &0.514967$\times 10^{-42}$  & 46.00 &0.140169$\times 10^{-40}$ &0.111034$\times 10^{-40}$ \\
   19.00 & 0.705006$\times 10^{-42}$ & 0.645647$\times 10^{-42}$ & 47.00 &0.149658$\times 10^{-40}$ &0.117888$\times 10^{-40}$ \\
   20.00 &0.869271$\times 10^{-42}$ &0.791903$\times 10^{-42}$ & 48.00 &0.159545$\times 10^{-40}$ &0.124970$\times 10^{-40}$ \\
   21.00 &0.105267$\times 10^{-41}$ & 0.953917$\times 10^{-42}$ &  49.00 &0.169835$\times 10^{-40}$ &0.132282$\times 10^{-40}$ \\ 
   22.00 &0.125572$\times 10^{-41}$ &0.113190$\times 10^{-41}$ & 50.00 &0.180532$\times 10^{-40}$ &0.139820$\times 10^{-40}$ \\
   23.00 &0.147902$\times 10^{-41}$ &0.132610$\times 10^{-41}$ & 51.00 &0.191641$\times 10^{-40}$ &0.147584$\times 10^{-40}$ \\
   24.00 &0.172324$\times 10^{-41}$ & 0.153684$\times 10^{-41}$ & 52.00 &0.203165$\times 10^{-40}$ &0.155572$\times 10^{-40}$ \\
   25.00 &0.198912$\times 10^{-41}$ &0.176455$\times 10^{-41}$ & 53.00 &0.215107$\times 10^{-40}$ &0.163783$\times 10^{-40}$ \\
   26.00 &0.227737$\times 10^{-41}$ &0.200959$\times 10^{-41}$ & 54.00 &0.227470$\times 10^{-40}$ &0.172212$\times 10^{-40}$ \\
   27.00 &0.258877$\times 10^{-41}$ &0.227231$\times 10^{-41}$ & 55.00 &0.240256$\times 10^{-40}$ &0.180857$\times 10^{-40}$ \\
   28.00 &0.292406$\times 10^{-41}$ &0.255312$\times 10^{-41}$ & 56.00 &0.253465$\times 10^{-40}$ &0.189715$\times 10^{-40}$ \\
   29.00 &0.328421$\times 10^{-41}$ &0.285245$\times 10^{-41}$ &  57.00 &0.267098$\times 10^{-40}$ &0.198781$\times 10^{-40}$ \\
   30.00 &0.366993$\times 10^{-41}$ &0.317070$\times 10^{-41}$ & 58.00 &0.281156$\times 10^{-40}$ &0.208052$\times 10^{-40}$ \\
   31.00 &0.408214$\times 10^{-41}$ &0.350828$\times 10^{-41}$ & 59.00 &0.295636$\times 10^{-40}$ & 0.217522$\times 10^{-40}$ \\
   32.00 &0.452166$\times 10^{-41}$ &0.386550$\times 10^{-41}$ & 60.00 &0.310538$\times 10^{-40}$ & 0.227187$\times 10^{-40}$ \\
   33.00 &0.498929$\times 10^{-41}$ &0.424272$\times 10^{-41}$ & & & \\ 
  \hline
  \end{tabular}
  \caption{Total cross sections for the neutral-current $\nu + ^{13}$C and $\overline{\nu} + ^{13}$C reactions with one neutron in the final state.}
  \label{nuandnubarNC}
\end{table}
These cross sections have several components which can help analysis of data from  supernova and neutron-spallation sources. These are compared in Fig. \ref{NCnemit}. 
In this figure 
the small difference between neutrino and antineutrino cross sections described above, comparable to the width of lines drawn, is ignored. The solid line refers to the total neutral-current one-neutron emission cross section. The dashed line refers to the cross section for transition to the  $0^+$ ground state in $^{12}$C:
\begin{equation}
\nu + ^{13}\text{C} \rightarrow \nu + n + ^{12}\text{C} ({\rm g.s.}), 
\end{equation}
and the dotted line refers to the cross section for transition to the  $2^+$ state in $^{12}$C at 4.44 MeV followed by decay into the $^{12}$C ground state: 
\begin{eqnarray}
\nu + ^{13}\text{C} \rightarrow \nu + n + ^{12}\text{C}(2^+) \nonumber \\
&\hookrightarrow&  ^{12}\text{C} ({\rm g.s.}) + \gamma (4.44 {\rm \> MeV}).  
\end{eqnarray}

Short-baseline reactor neutrino experiments identified a shape distortion in the 5-7 MeV range in the measured neutrino spectrum \cite{Balantekin:2016vjt}. This shape distortion appears as an excess over the predicted spectra. A recent proposal suggests \cite{Berryman:2018jxt} a beyond the Standard Model solution to resolve this issue: Non-standard neutrino interactions which induce the reaction $^{13}$C( $\overline{\nu}, \overline{\nu}'$ n)$^{12}$C$^*$ followed by the de-excitation of $^{12}$C$^*$ yielding a prompt 4.44 MeV photon. The neutron produced would then lose its kinetic energy scattering off protons yielding scintillation light. It was proposed that this scintillation light along with the prompt photon would mimic the spectral distortion around 5 MeV. To help assess further investigation of such processes we tabulate the Standard Model $^{13}$C( $\overline{\nu}, \overline{\nu}'$ n)$^{12}$C$^*$ neutrino cross sections 
in Table \ref{4.4MeVcs}. 

\begin{table}[b]
  \centering
  \begin{tabular}{@{} cccccc @{}}
    \hline
     E (MeV) & $\sigma (\nu)$ (cm$^2$) &  $\sigma (\overline{\nu})$ (cm$^2$) &   E (MeV) & $\sigma (\nu) $ (cm$^2$) &  $\sigma (\overline{\nu}$) (cm$^2$) \\
    \hline 
    8.00 & 5.53038 $\times 10^{-57}$ & 0 &                                                   18.00 & 5.76896 $\times 10^{-44}$ & 5.41018 $\times 10^{-44}$   \\
    9.00  &1.15214 $\times 10^{-53}$ & 1.03756 $ \times 10^{-53}$   &         19.00&  7.64158 $\times 10^{-44}$ & 7.13818 $\times 10^{-44}$ \\
   10.00 & 6.12058 $\times 10^{-48}$ & 5.95485 $\times 10^{-48}$   &         20.00 & 9. 84805 $\times 10^{-44}$ & 9.16268 $\times 10^{-44}$ \\
   11.00 &  7.68861$ \times 10^{-46}$ & 7.42909 $\times 10^{-46}$   &     21.00  & 1.24147 $\times 10^{-43}$ & 1.15038 $\times 10^{-43}$ \\
   12.00 & 2.81923 $\times 10^{-45}$ & 2.70849 $\times 10^{-45}$   &      22.00 & 1.53709 $\times 10^{-43}$ & 1.41834 $\times 10^{-43}$   \\ 
   13.00 & 6.26649 $\times 10^{-45}$ & 5.98553 $\times 10^{-45}$   &     23.00  & 1.87481 $ \times 10^{-43}$ & 1.72247 $ \times 10^{-43}$ \\
   14.00 &  1.16558 $\times 10^{-44}$ & 1.10949 $\times 10^{-44} $  &   24.00 & 2.25800 $ \times 10^{-43}$ & 2.06520 $\times 10^{-43}$    \\
   15.00 &  1.92811 $\times 10^{-44}$ & 1.82982 $ \times 10^{-44}$  &     25.00 & 2.69019 $\times 10^{-43}$ & 2.44904 $\times 10^{-43}$ \\ 
    16.00&  2.93457 $ \times 10^{-44}$&  2.77418 $\times 10^{-44}$  &     26.00 & 3.17505 $\times 10^{-43}$ & 2.87653 $\times 10^{-43}$ \\ 
    17.00 & 4.20674 $ \times 10^{-44}$ & 3.96067 $\times 10^{-44}$ &     27.00  & 3.71642 $\times 10^{-43}$ &  3.35024 $\times  10^{-43}$ \\
    \hline
  \end{tabular}
  \caption{$^{13}$C ($\nu, \nu'$ n)$^{12}$C$^*$ and $^{13}$C( $\overline{\nu}, \overline{\nu}'$ n)$^{12}$C$^*$ neutrino cross sections leading to the 4.44 MeV state in $^{12}$C calculated using only the Standard Model interactions.}
  \label{4.4MeVcs}
\end{table}

\begin{figure}[t]
\begin{center}
\includegraphics[scale=0.5]{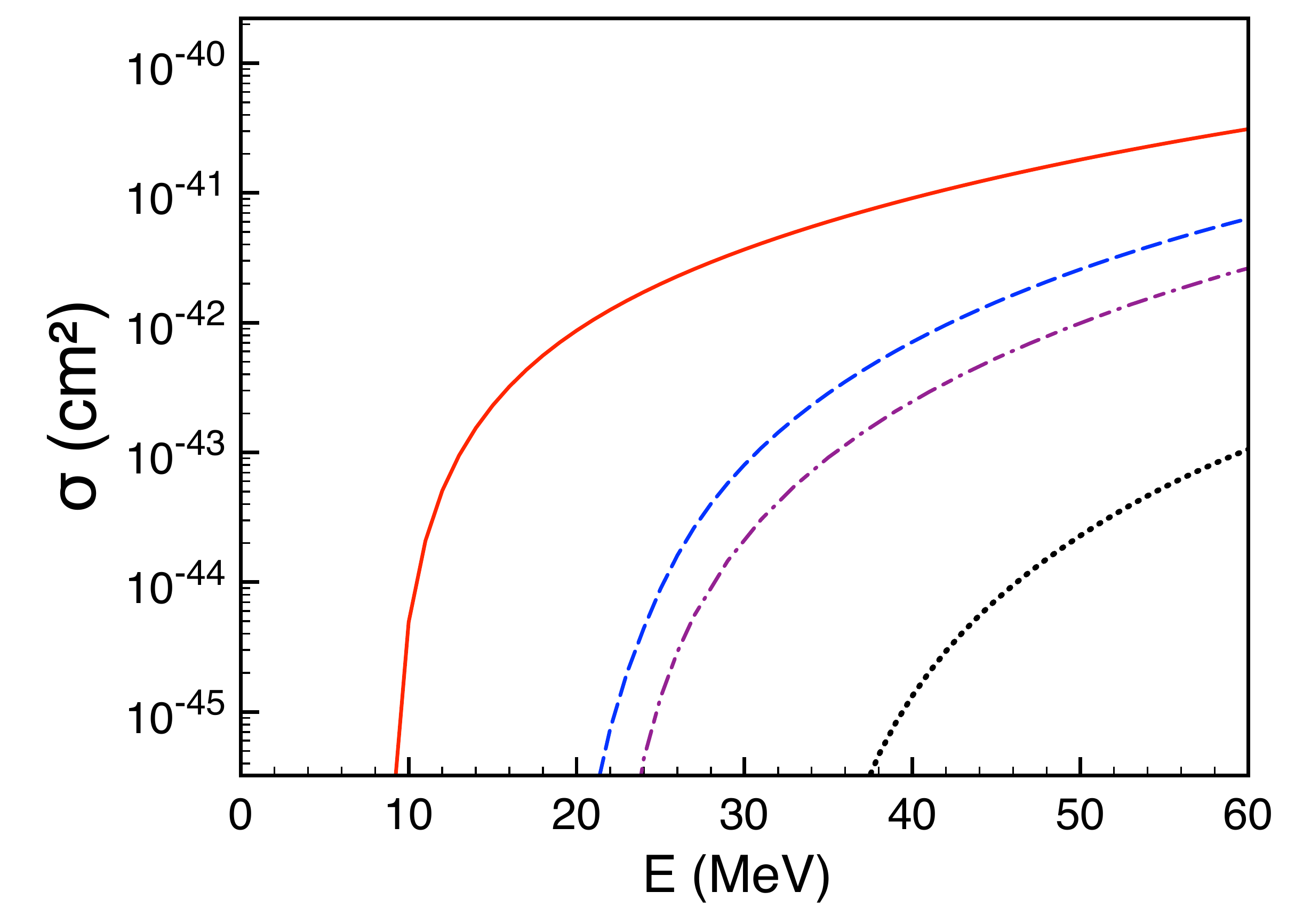}
\caption{A comparison of the neutron emission cross sections for $\nu_e$ interacting with carbon isotopes. Neutral-current contribution to 
neutron emission in the reaction $\nu + ^{13}$C (solid line),  neutral-current contribution to neutron emission in  the reaction $\nu_e + ^{12}$C (dashed line), charged-current contribution to neutron emission in the reaction $\nu + ^{13}$C (dot-dashed line), and charged-current contribution to neutron emission in the reaction $\nu_e + ^{12}$C (dotted line). The $^{12}$C data is taken from Ref. \cite{Yoshida:2008zb}.}
\label{C_neutron_comp}
\end{center}
\end{figure}

Finally a comparison of the various neutron emission cross sections for $\nu_e$ interacting with carbon isotopes is given in Figure \ref{C_neutron_comp}. 
The cross sections are roughly proportional to the phase space of the final states, $(E_{\nu}-E_{th})^2$, where $E_{\nu}$ is the initial neutrino energy and $E_{th}$ is the threshold energy for neutron emissions measured from the ground state of the parent nucleus. 
For neutral-current reactions, $E_{th}$ is equal to the neutron separation energy ($S_n$), and $E_{th}$ =4.946 MeV and 18.721 MeV for $^{13}$C and $^{12}$C, respectively. For charged-current reactions, $E_{th}$ is the sum of $S_n$ in the daughter nucleus and the $\beta^{+}$-decay Q-value. They are $E_{th}$ =22.284 MeV ($S_n$ =20.064 MeV, Q =2.220 MeV) and 32.378 MeV ($S_n$ =15.040 MeV, Q =17.338 MeV) for $^{13}$C and $^{12}$C, respectively.
The difference in the magnitude of the calculated cross sections in Figure \ref{C_neutron_comp} can be understood in terms of the difference in the threshold energies except for some details due to the difference between $^{12}$C and $^{13}$C.  
In case of neutral-current reaction on $^{13}$C, the Gamow-Teller transition to the 1/2$^{-}$ state at $E_x$ =8.860 MeV gives the dominant contribution to the cross section near threshold, that is, at $E_{\nu}$ =9-12 MeV. At $E_{\nu} \geq$ 15 MeV, the Gamow-Teller transition to the 3/2$^{-}$ state at $E_x$ =9.898 MeV gives a contribution comparable to the 1/2$^{-}$ state.    
In case of neutral-current reaction on $^{12}$C, excitations of spin-dipole states (0$^{-}$, 1$^{-}$ and 2$^{-}$ states) above the threshold energy give more important contributions to the neutron-emission cross section than those of 1$^{+}$ states. Note that the 1$^{+}$ state at $E_x$ =15.11 MeV with the largest magnetic dipole strength is below the neutron emission threshold.    

\section{Conclusions}

Since $^{13}$C is naturally present in carbon-based scintillators, it is important to know its interaction cross section with neutrinos to achieve high precision in neutrino experiments using such scintillators. In an earlier publication we presented calculations of those cross sections at reactor neutrino energies. In this paper these cross sections both for charged-current and 
neutral-current reactions are given at higher energies, 
relevant to terrestrial supernova observations and measurements at the spallation neutron sources (such as the Oak Ridge facility used in the experiment of Ref. 
\cite{Akimov:2017ade}). Since there are ongoing and planned neutrino coherent scattering experiments at those 
spallation neutron sources, those cross sections are also provided. Since the processes with neutrons are present in the final state and could provide a convenient signal via the 2.2 MeV photons as well as help in 
the searches for physics beyond the Standard Model we provided the Standard Model values of the cross section for those processes as well. 

%%%%%%%%%%%%%%%%%%%%%%%%%%%%%%%%%%%%%%
\section*{Acknowledgments} 
%%%%%%%%%%%%%%%%%%%%%%%%%%%%%%%%%%%%%%

This work was supported in part by Grants-in-Aid for Scientific Research of JSPS (15H03665, 15K05090, 17K05459), 
in part by the U.S. National Science Foundation Grants No. PHY-1806368 and and PHY-1630782. 
ABB is also supported in part by the visiting professor program at the National Astronomical Observatory of Japan.

%%%%%%%%%%%%%%%%%%%%%%%%
%%%%%%%%%%%%%%%%%%%


\begin{thebibliography}{99}
%%%%%%%%%%%%%%%%%%%

\bibitem{Suzuki:2003}
T.~Suzuki, R.~Fujimoto and T.~Otsuka, 
 Phys.\ Rev.\ C {\bf 67}, 044302 (2003).
  
%\cite{Adey:2018zwh}
\bibitem{Adey:2018zwh} 
  D.~Adey {\it et al.} [Daya Bay Collaboration],
  %``Measurement of the Electron Antineutrino Oscillation with 1958 Days of Operation at Daya Bay,''
  Phys.\ Rev.\ Lett.\  {\bf 121}, no. 24, 241805 (2018)
  doi:10.1103/PhysRevLett.121.241805
  [arXiv:1809.02261 [hep-ex]].
  %%CITATION = doi:10.1103/PhysRevLett.121.241805;%%  
  
%\cite{Bak:2018ydk}
\bibitem{Bak:2018ydk} 
  G.~Bak {\it et al.} [RENO Collaboration],
  %``Measurement of Reactor Antineutrino Oscillation Amplitude and Frequency at RENO,''
  Phys.\ Rev.\ Lett.\  {\bf 121}, no. 20, 201801 (2018)
  doi:10.1103/PhysRevLett.121.201801
  [arXiv:1806.00248 [hep-ex]].
  %%CITATION = doi:10.1103/PhysRevLett.121.201801;%%  
  
%\cite{Kaneda:2018nma}
\bibitem{Kaneda:2018nma} 
  M.~Kaneda [Double Chooz Collaboration],
  %``Recent Results from Double Chooz,''
  Phys.\ Part.\ Nucl.\  {\bf 49}, no. 4, 709 (2018).
  doi:10.1134/S1063779618040317
  %%CITATION = doi:10.1134/S1063779618040317;%%  
  
%\cite{Agostini:2018uly}
\bibitem{Agostini:2018uly} 
  M.~Agostini {\it et al.} [BOREXINO Collaboration],
  %``Comprehensive measurement of $pp$-chain solar neutrinos,''
  Nature {\bf 562}, no. 7728, 505 (2018).
  doi:10.1038/s41586-018-0624-y
  %%CITATION = doi:10.1038/s41586-018-0624-y;%%  
  
  %\cite{Miramonti:2018noj}
\bibitem{Miramonti:2018noj} 
  L.~Miramonti [JUNO Collaboration],
  %``Neutrino Physics and Astrophysics with the JUNO Detector,''
  Universe {\bf 4}, no. 11, 126 (2018).
  doi:10.3390/universe4110126
  %%CITATION = doi:10.3390/universe4110126;%%
  
 %\cite{Wurm:2011zn}
\bibitem{Wurm:2011zn} 
  M.~Wurm {\it et al.} [LENA Collaboration],
  %``The next-generation liquid-scintillator neutrino observatory LENA,''
  Astropart.\ Phys.\  {\bf 35}, 685 (2012)
  doi:10.1016/j.astropartphys.2012.02.011
  [arXiv:1104.5620 [astro-ph.IM]].
  %%CITATION = doi:10.1016/j.astropartphys.2012.02.011;%% 

%\cite{Athanassopoulos:1997rm}
\bibitem{Athanassopoulos:1997rm} 
  C.~Athanassopoulos {\it et al.} [LSND Collaboration],
  %``Measurements of the reactions C-12 (electron-neutrino, e-) N-12 (g.s.) and C-12 (electron-neutrino, e-) N*-12,''
  Phys.\ Rev.\ C {\bf 55}, 2078 (1997)
  doi:10.1103/PhysRevC.55.2078
  [nucl-ex/9705001].
  %%CITATION = doi:10.1103/PhysRevC.55.2078;%%  
  
 %\cite{Arafune:1988hx}
\bibitem{Arafune:1988hx}   
  J.~Arafune, M.~Fukugita, Y.~Kohyama and K.~Kubodera,
  %``C-13 As A Solar Neutrino Detector,''
  Phys.\ Lett.\ B {\bf 217}, 186 (1989).
  %%CITATION = PHLTA,B217,186;%%

%\cite{Fukugita:1989wv}
\bibitem{Fukugita:1989wv} 
  M.~Fukugita, Y.~Kohyama, K.~Kubodera and T.~Kuramoto,
  %``REACTION CROSS-SECTIONS FOR neutrino C-13 ---> $\times 10^{- N-13 AND neutrino C-13 ---> neutrino-prime C*13 FOR LOW-ENERGY NEUTRINOS,''
  Phys.\ Rev.\ C {\bf 41}, 1359 (1990).
  %%CITATION = PHRVA,C41,1359;%%
  
%\cite{Cohen:1965qa}
\bibitem{Cohen:1965qa} 
  S.~Cohen and D.~Kurath,
  %``Effective interactions for the 1p shell,''
  Nucl.\ Phys.\  {\bf 73}, 1 (1965).
  %%CITATION = NUPHA,73,1;%%  
  
%\cite{Millener:1975zz}
\bibitem{Millener:1975zz} 
  D.~J.~Millener and D.~Kurath,
  %``The particl$\times 10^{-hole interaction and the beta decay of B-14,''
  Nucl.\ Phys.\ A {\bf 255}, 315 (1975).
  %%CITATION = NUPHA,A255,315;%%  
 

%\cite{Arima:1988xa}
\bibitem{Arima:1988xa} 
  A.~Arima, K.~Shimizu, W.~Bentz and H.~Hyuga,
  %``Nuclear Magnetic Properties And Gamow-teller Transitions,''
  Adv.\ Nucl.\ Phys.\  {\bf 18}, 1 (1987).
  %%CITATION = ANUPB,18,1;%%
  
%\cite{Pourkaviani:1991pb}
\bibitem{Pourkaviani:1991pb} 
  M.~Pourkaviani and S.~L.~Mintz,
  %``Neutrino reaction in C-13,''
  J.\ Phys.\ G G {\bf 17}, 1139 (1991).
  %%CITATION = JPHGB,G17,1139;%%  
  
  %\cite{Mintz:2000xv}
\bibitem{Mintz:2000xv} 
  S.~L.~Mintz,
  %``Inclusive neutrino reactions in C-13,''
  Nucl.\ Phys.\ A {\bf 672}, 503 (2000).
  %%CITATION = NUPHA,A672,503;%%

%\cite{Suzuki:2012aa}
\bibitem{Suzuki:2012aa} 
  T.~Suzuki, A.~B.~Balantekin and T.~Kajino,
  %``Neutrino Capture on $^{13}$C,''
  Phys.\ Rev.\ C {\bf 86}, 015502 (2012)
  [arXiv:1204.4231 [nucl-th]].
  %%CITATION = ARXIV:1204.4231;%%

%\cite{Scholberg:2012id}
\bibitem{Scholberg:2012id} 
  K.~Scholberg,
  %``Supernova Neutrino Detection,''
  Ann.\ Rev.\ Nucl.\ Part.\ Sci.\  {\bf 62}, 81 (2012)
  [arXiv:1205.6003 [astro-ph.IM]].
  %%CITATION = ARXIV:1205.6003;%%

%\cite{Dasgupta:2011wg}
\bibitem{Dasgupta:2011wg} 
  B.~Dasgupta and J.~.F.~Beacom,
  %``Reconstruction of supernova $\nu_\mu$, $\nu_\tau$, anti-$\nu_\mu$, and anti-$\nu_\tau$ neutrino spectra at scintillator detectors,''
  Phys.\ Rev.\ D {\bf 83}, 113006 (2011)
  [arXiv:1103.2768 [hep-ph]].
  %%CITATION = ARXIV:1103.2768;%%

%\cite{Kuo:1967}
\bibitem{Kuo:1967}
T.~T.~S. Kuo, Nucl.\ Phys.\ A {\bf 103}, 71 (1967).

%\cite{Suzuki:2006qd}
\bibitem{Suzuki:2006qd} 
  T.~Suzuki, S.~Chiba, T.~Yoshida, T.~Kajino and T.~Otsuka,
  %``Neutrino nucleus reactions based on new shell model Hamiltonians,''
  Phys.\ Rev.\ C {\bf 74}, 034307 (2006)
  [nucl-th/0608056].
  %%CITATION = NUCL-TH/0608056;%%

%\cite{Otsuka:2005zz}
\bibitem{Otsuka:2005zz} 
  T.~Otsuka, T.~Suzuki, R.~Fujimoto, H.~Grawe and Y.~Akaishi,
  %``Evolution of Nuclear Shells due to the Tensor Force,''
  Phys.\ Rev.\ Lett.\  {\bf 95}, 232502 (2005).
  %%CITATION = PRLTA,95,232502;%%

%cite{Suzuki:2013}
\bibitem{Suzuki:2013}
  T.~Suzuki and T.~Kajino,
  J.\ Phys.\ G {\bf 40}, 083101 (2013).
%

\bibitem{Walecka:1975}
J.~D.~Walecka, in {\it Muon Physics}, edited by V.~H.~Highes 
and C.~S.~Wu (Academic, New York, 1975), Vol. II;\\
J.~S.~O'Connell, T.~W.~Donnelly and J.~D.~Walecka,
Phys.\ Rev.\ C {\bf 6}, 719 (1972);\\
T.~W.~Donnelly and J.~D.~Walecka, Nucl.\ Phys.\ {\bf A274}, 368 (1976);\\
T.~W.~Donnely and W.~C.~Haxton, Atomic Data Nucl. Data Tables 
{\bf 23}, 103 (1979). 

%cite{Walter:1986}
\bibitem{Walter:1986}
R.~L.~Walter and P.~P.~Guss, in {\it Proceedings of the International
Conference on Nuclear Data for Basic and Applied Science},
edited by P.~G.~Young, Vol. 2 (Gordon and Breach, New Yprk, 1986), p. 1079.
%Santa Fe, May 13-17, 1985, p. 1079 (unpublished).

%cite{Avrigeanu:1994}
\bibitem{Avrigeanu:1994}
V.~Avrigeanu, P.~E.~Hodgson, and M.~Avrigeanu, Phys.\ Rev.\ C
{\bf 49}, 2136 (1994).
%

%cite{Belgya:2006}
\bibitem{Belgya:2006}
T.~Belgya, O.~Bersillon, R.~Capote, T.~Fukahori, G.~Zhigang,
S.~Goriely, M.~Herman, A.~V.~Ignatyuk, S.~Kailas, A.~Koning,
P.~Oblozhinsky, V.~Plujko, and P.~Young, {\it Handbook for Calculations
of Nuclear Reaction Data: Reference Input Parameter
Library} (IAEA, Vienna, 2006). http://wwwnds.iaea.org/RIPL-2/
%

%\cite{Ianni:2005ki}
\bibitem{Ianni:2005ki} 
  A.~Ianni, D.~Montanino and F.~L.~Villante,
  %``How to observe B-8 solar neutrinos in liquid scintillator detectors,''
  Phys.\ Lett.\ B {\bf 627}, 38 (2005)
  doi:10.1016/j.physletb.2005.08.122
  [physics/0506171 [physics.ins-det]].
  %%CITATION = doi:10.1016/j.physletb.2005.08.122;%%
  
%\cite{Bolozdynya:2012xv}
\bibitem{Bolozdynya:2012xv} 
  A.~Bolozdynya {\it et al.},
  %``Opportunities for Neutrino Physics at the Spallation Neutron Source: A White Paper,''
  arXiv:1211.5199 [hep-ex].
  %%CITATION = ARXIV:1211.5199;%%  

%\cite{Freedman:1973yd}
\bibitem{Freedman:1973yd} 
  D.~Z.~Freedman,
  %``Coherent neutrino nucleus scattering as a probe of the weak neutral current,''
  Phys.\ Rev.\ D {\bf 9}, 1389 (1974).
  doi:10.1103/PhysRevD.9.1389
  %%CITATION = doi:10.1103/PhysRevD.9.1389;%%

%\cite{Freedman:1977xn}
\bibitem{Freedman:1977xn} 
  D.~Z.~Freedman, D.~N.~Schramm and D.~L.~Tubbs,
  %``The Weak Neutral Current and Its Effects in Stellar Collapse,''
  Ann.\ Rev.\ Nucl.\ Part.\ Sci.\  {\bf 27}, 167 (1977).
  doi:10.1146/annurev.ns.27.120177.001123
  %%CITATION = doi:10.1146/annurev.ns.27.120177.001123;%%

%\cite{Drukier:1983gj}
\bibitem{Drukier:1983gj} 
  A.~Drukier and L.~Stodolsky,
  %``Principles and Applications of a Neutral Current Detector for Neutrino Physics and Astronomy,''
  Phys.\ Rev.\ D {\bf 30}, 2295 (1984).
  doi:10.1103/PhysRevD.30.2295
  %%CITATION = doi:10.1103/PhysRevD.30.2295;%%

%\cite{Akimov:2017ade}
\bibitem{Akimov:2017ade} 
  D.~Akimov {\it et al.} [COHERENT Collaboration],
  %``Observation of Coherent Elastic Neutrino-Nucleus Scattering,''
  Science {\bf 357}, no. 6356, 1123 (2017)
  doi:10.1126/science.aao0990
  [arXiv:1708.01294 [nucl-ex]].
  %%CITATION = doi:10.1126/science.aao0990;%%

 %\cite{Yoshida:2008zb}
\bibitem{Yoshida:2008zb} 
  T.~Yoshida, T.~Suzuki, S.~Chiba, T.~Kajino, H.~Yokomakura, K.~Kimura, A.~Takamura and D.~H.~Hartmann,
  %``Neutrino-Nucleus Reaction Cross Sections for Light Element Synthesis in Supernova Explosions,''
  Astrophys.\ J.\  {\bf 686}, 448 (2008)
  doi:10.1086/591266
  [arXiv:0807.2723 [astro-ph]].
  %%CITATION = doi:10.1086/591266;%% 

%\cite{Patton:2012jr}
\bibitem{Patton:2012jr} 
  K.~Patton, J.~Engel, G.~C.~McLaughlin and N.~Schunck,
  %``Neutrino-nucleus coherent scattering as a probe of neutron density distributions,''
  Phys.\ Rev.\ C {\bf 86}, 024612 (2012)
  doi:10.1103/PhysRevC.86.024612
  [arXiv:1207.0693 [nucl-th]].
  %%CITATION = doi:10.1103/PhysRevC.86.024612;%%

%\cite{Balantekin:2016vjt}
\bibitem{Balantekin:2016vjt} 
  A.~B.~Balantekin,
  %``Reactor antineutrinos and nuclear physics,''
  Eur.\ Phys.\ J.\ A {\bf 52}, no. 11, 341 (2016).
  doi:10.1140/epja/i2016-16341-5
  %%CITATION = doi:10.1140/epja/i2016-16341-5;%%  

%\cite{Berryman:2018jxt}
\bibitem{Berryman:2018jxt} 
  J.~M.~Berryman, V.~Brdar and P.~Huber,
  %``Particle physics origin of the 5 MeV bump in the reactor antineutrino spectrum?,''
  Phys.\ Rev.\ D {\bf 99}, no. 5, 055045 (2019)
  doi:10.1103/PhysRevD.99.055045
  [arXiv:1803.08506 [hep-ph]].
  %%CITATION = doi:10.1103/PhysRevD.99.055045;%%

\end{thebibliography}
\end{document}